\documentclass[reprint,aps,pra,twocolumn]{revtex4-2}

\usepackage{graphicx}
\usepackage{dcolumn}
\usepackage{bm}
\usepackage{hyperref}
\usepackage[mathlines]{lineno}

\hypersetup{colorlinks=true, citecolor=blue, urlcolor=blue, linkcolor=blue}
\usepackage{slashed}  
\usepackage{amsmath}
\usepackage{amssymb} 
\usepackage{natbib}
\usepackage{mathrsfs}
\usepackage{flushend}
\makeatletter

\newcommand{\Rmnum}[1]{\expandafter\@slowromancap\romannumeral #1@}
\makeatother
\allowdisplaybreaks[3]

\begin{document}
\preprint{APS/123-QED}

\title{Sub-Rayleigh Ghost Imaging via Structured Speckle Illumination}


\author{Liming Li}
\email{liliming@sdut.edu.cn}
\affiliation{School of Physics and Optoelectronic Engineering, Shandong University of Technology, Zibo 255049, China\\}

\date{\today}

\begin{abstract}
The structured illumination is adopted widely in the super-resolution microscopy imaging. Here, we studied the ghost imaging scheme with sinusoidal structured speckle illumination, whose spatial resolution can surpass the Rayleigh-resolution limit by a factor of 2. In addition, ghost image with a higher spatial resolution can be realized by a pseudo-modulation structured illumination, whose modulation frequency breaks the upper limit of imaging system. Finally, our research verifies that the structural speckle illumination, a free-standing sub-Rayleigh technology, can be grafted with a variety of sub-Rayleigh ghost imaging schemes to further improve the spatial resolution of ghost image, such as the pseudo-inverse ghost imaging.
\end{abstract}
\maketitle
\section{Introduction}
Due to wave property and finite diffraction aperture, an incoherence imaging system is of limited imaging resolution, known as the Rayleigh-resolution limit $\Delta x_R=0.61 \lambda/\text{NA}$~\cite{Rayleigh1897Investigations}, where $\lambda$ is the wavelength of light and NA is the numerical aperture of imaging system~\cite{Goodman2005Introduction2}. To break through this limit, Heintzmann and Cremer in 1999 proposed a super-resolution imaging scheme, that the object was illuminated by a series of structured light field created by grating diffraction with coherent light~\cite{Heintzmann1998Laterally}. The detected spectrum of the imaging system can be shifted. Here, the structured illumination is a series of sinusoidal patterns. Thus, the structured illumination as a super-resolution technology is based on the perspective of the extended spectrum~\cite{Gustafsson2001Microscopy, Frohn2000PNAS}. Because of fast imaging speed and high photon efficiency, this super-resolution technology has been applied to a wide variety of biological problems, which gives birth to a hot academic topic the structured illumination microscopy (SIM)~\cite{Heintzmann2017Super, Strohl2016Frontiers}. Moreover, due to independence and simplicity, SIM is compatible widely with other super-resolution schemes result to many hybrid super-resolution imaging schemes~\cite{Xue2018Three, Wang2023Hybrid, Chen2023Superresolution}.

Comparing with the structured illumination imaging, the ghost imaging (GI) has also rely on light field irradiating the test object. As a promising imaging technology, GI is originated from two-arms light intensity correlation between a point-signal arm with a non-spatial resolution bucket (or single-pixel) detector which collects transmitted light from the object, and a array-signal arm with a pixelated array of detector which never interacts with the object and only records the light field distribution on the mirror plane of the object. Actually, those mirror planes, where dynamic light field has exactly the same temporal and spatial distribution, can be created either classical sources~\cite{Bennink2002CoincidencePRL, Valencia2005TwoPRL, Ferri2005HighPRL, ChenHigh-visibility2010, Wu2005Correlated, Liu2014Lensless} with the help of a beam splitter (BS) or quantum sources~\cite{Klyshko1988Two, Pittman1995PRA, Abouraddy2004Entangled, Aspden2015Photon}. At first, people believed that quantum sources can only realize GI, but after a while the research of GI with classical source became popular, such as pseudo-thermal light~\cite{Valencia2005TwoPRL, Ferri2005HighPRL, ChenHigh-visibility2010}, true thermal light~\cite{Wu2005Correlated}, and even sun light~\cite{Liu2014Lensless}, etc. Remarkably, the GI scheme with a priori light source can be change into the single-arm computational ghost imaging (CGI)~\cite{Shapiro2008Computational} because the array-signal can be acquired via the diffraction integral formula~\cite{Goodman2005Introduction2}. Interestingly, this simplified single-arm GI scheme has gradually converted into to a novel research topic the single-pixel imaging (SPI)~\cite{	Bromberg2009Ghost,Sun2013science,XU20181000,Edgar2018Principles,Sun2019Single,Hong2019Sub}, which brings many advantages such as free of aberration caused by the multi-wavelength dispersion, wide spectral range detection, and higher detection signal noise ratio in weak light environment. As one of key device in computational imaging, the spatial light modulator (SLM) can not only simplify the experimental device from two-arms GI to single-arm CGI~\cite{Shapiro2008Computational,Bromberg2009Ghost,Hong2019Sub,Li2019PRA,Li2021Eigenmode}, but be widely used in structured light imaging~\cite{Kner2009Super, Hirvonen2009Structured, Chang2009Isotropic, Guo2018Visualizing}. Consequently, a hybrid imaging scheme between CGI and structured illumination imaging is possible theoretically to improve the spatial resolution of ghost image with the help of SLM.

Similar to traditional incoherent lens imaging, the spatial resolution of thermal GI also follows the Rayleigh-resolution limit~\cite{Ferri2005HighPRL, ChenHigh-visibility2010, Li2019PRA}. To improve the spatial resolution of classical light GI, some methods have been studied, such as compressed sensing~\cite{Gong2015Highsr,Lochocki2021Ultimate}, pseudo-inverse ghost imaging (PGI)~\cite{Zhang2014Object,Gong2015Highpr}, non-Rayleigh speckle~\cite{Zhang2016High,KYRUS2016High}, speckle spatial filtering~\cite{Sprigg2016Super,Meng2018Super}, optical fluctuation~\cite{Hong2019Sub}, and deep learning~\cite{Wang2022Far} etc. In this work, we reported a far-field sub-Rayleigh structured illumination ghost imaging (SIGI) scheme, which clearly show the enhancement of spatial resolution by a wider detected spectrum domain. Here, the structured illumination is sinusoidal speckle, which can be obtained by holographic projection or the pseudo-modulation method. Comparing with the holographic projection a limited modulation frequency leading to a factor of 2 enhancement on the spatial resolution, the pseudo-modulation is unconstrained on the modulation frequency. In addition, the diffractive SIGI scheme was also analyzed when the test object and the bucket detector are not close contact. Finally, a grafted imaging scheme between two sub-Rayleigh GI schemes (SIGI and PGI) is discussed in detail.

\section{Theory}\label{Theory}
\begin{figure}[b]
	\centering
	\includegraphics[width=0.45\textwidth]{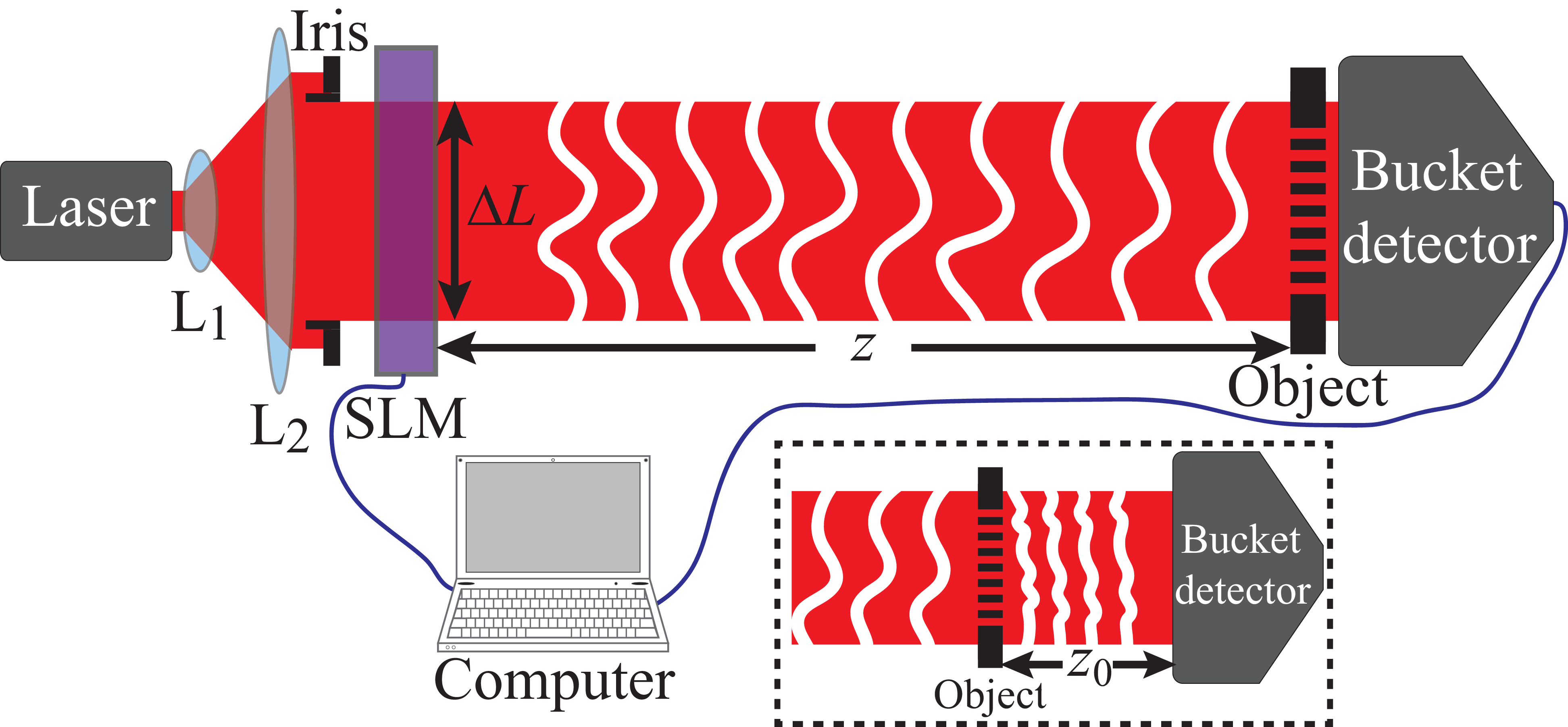}
	\caption{Schematic diagram of single-pixel sub-Rayleigh SIGI. $\text{L}_1$, $\text{L}_2$: lenses; SLM: spatial light modulator. The inset at the right-bottom corner shows the diffractive SIGI scheme with distance $z_0$ between the object plane and the bucket detector.} \label{SIGIdiagram}
\end{figure}
To demonstrate the high-resolution of SIGI, figure~\ref{SIGIdiagram} depicts the numerical experimental setup, which is identical with the CGI~\cite{Shapiro2008Computational}. A continuous-wave $\lambda$ = 532 nm single-mode laser beam was expanded and collimated by two lenses $L_1$ and $L_2$, and then passed through an iris and a phase-only transmission SLM (an element pixel size of 20 × 20 $\mu \text{m}^{2}$ and a total pixels 512 $\times$ 512). The iris was placed as close as possible to the SLM, and its diameter $\Delta L$ was set to be 1.2 mm. The computer-controlled SLM is used to modulate the wavefront of the laser beam to create the source field $E(r_\text{s},t)\propto  e^{i\theta (r_\text{s},t)}$, whose phasor $\theta (r_\text{s},t)$ is dynamic prepared in advance. Here, the effective aperture of $E(r_\text{s},t)$ is controlled by the iris with a diameter $\Delta L$. The source field $E(r_\text{s},t)$ undergos quasi monochromatic paraxial diffraction along the optical axe, over an $z$ free-space path, yielding the structured illumination field $E(r_\text{o},t)$ on the object's plane. To create far-field GI scheme, a lens phase factor $e^{-ikr_{\text{s}}^{2}/\left ( 2z \right )} $ is superposed on the phasor $\theta (r_\text{s},t)$ of source field, where $k=2\pi/\lambda$ is the wave number and $z$ = 0.8 $\text{m}$ is the distance between the object and the SLM. An amplitude transmission object $O(r_\text{o})$, located immediately in front of a single-pixel bucket photodetector, are illuminated by the structured field $E(r_\text{o},t)$.

Here, the traditional CGI is given by~\cite{Katz2009Compressive,Gong2015Highpr,Li2019PRA}
\begin{equation}\label{C01}
	\begin{split}
		GI{(r_\text{o})} = \frac{{\text{1}}}{{{N}}}\sum\limits_{j = 1}^{N} {  \left(  {I_j}{(r_\text{o})}-\left\langle {I{(r_\text{o})}} \right\rangle  \right)   {BI_{j}  } }\, ,
	\end{split}
\end{equation}
where $N$ is the number of total frames, ${I_j}{(r_\text{o})}=\left |E(r_\text{o},t)\right |^{2}$ is the $j$-th frame array-signal namely the instantaneous intensity distribution on the object plane at $t$ time and $\left\langle {I{(r_\text{o})}} \right\rangle=\frac{1}{N}  {\textstyle \sum_{j=1}^{N}} I_{j} \left (r_\text{o}  \right ) $ is the ensemble average of all instantaneous intensity, $BI_{j}=\int{I_j}{(r_\text{o})}O(r_\text{o})dr_\text{o}$ is the bucket signal which is the total light intensity passing through the object. Here, the $j$-th frame source phase $\theta_j (r_\text{s},t)$ prepared in advance will be loaded experimentally at time sequence $t$. Therefore, illuminated with the priori source field $E(r_\text{s},t)$, a SPI setup can be created by a single-pixel bucket detector which collects those $N$ transmission coefficient $BI_{j}$ to reconstruct the spatial distribution of the object~\cite{Shapiro2008Computational}. For a traditional far-field CGI, the numerical aperture $\text{NA}=\frac{\Delta L}{2z}$ is decided by the diameter $\Delta L$ of thermal source (or iris) and the diffraction distance $z$ between the source and the test object.

The matrix operations offer a concise way to express the correlation algorithm of GI. Provided that the speckle ${I_j}{(r_\text{o})}$ is composed of $m\times n$ pixels so that the speckle sequence ${I_j}(r_\text{o})$
($j=1, 2,\cdots, N$) creates the $N\times K$ measurement matrix $\Phi_{(N\times K)}$ ($K=m \times n$), namely
\begin{equation}\label{C02}
	\Phi  = \left[ \begin{array}{l}
		{I_1}\left( {1,1} \right),{\kern 1pt} {I_1}\left( {1,2} \right),{\kern 1pt}  \cdots ,{\kern 1pt} {I_1}\left( {m,n} \right)\\
		{I_2}\left( {1,1} \right),{\kern 1pt} {I_2}\left( {1,2} \right),{\kern 1pt}  \cdots ,{\kern 1pt} {I_2}\left( {m,n} \right)\\
		\quad\vdots \quad\quad\quad\quad\quad\ddots \quad\quad\quad\quad\vdots \\
		{I_N}\left( {1,1} \right),{\kern 1pt} {I_N}\left( {1,2} \right),{\kern 1pt}  \cdots ,{\kern 1pt} {I_N}\left( {m,n} \right)
	\end{array} \right].
\end{equation}
Further, a matrix $\Psi_{(N\times K)}  = \Phi_{(N\times K)}  - {\text{E}_{(N\times 1)}}\left\langle \Phi  \right\rangle_{(1\times K)} $ is defined, where ${\text{E}}_{(N\times 1)}$ is a $N\times 1$ column vector with all elements 1 and $\left\langle \Phi  \right\rangle_{(1\times K)} {\rm{ = }}\frac{1}{N}  {\textstyle \sum_{j=1}^{N}} {\Phi_{j{(1\times K)}}} $ is the average of the measurement matrix in column direction. Here, ${\Phi_{j{(1\times K)}}}$ denotes the $j$-th row vector of the matrix $\Phi_{(N\times K)}$. Therefore, the matrix format of Eq.~(\ref{C01}) can be simply expressed as
\begin{equation}\label{C03}
	GI = \frac{1}{N} {\Psi^\text{T}}\Phi O,
\end{equation}
where $\Psi^\text{T}$ denotes the transposition of the matrix $\Psi$ and $O = {\left[ {o\left( {1,1} \right),{\kern 1pt} o\left( {1,2} \right),{\kern 1pt}  \cdots ,{\kern 1pt} o\left( {m,n} \right)} \right]^{\rm{T}}}$ is the $K\times 1$ column vector which originates from the test object $O(r_\text{o})$.

To realize the SIGI scheme, the illumination speckle are modulated by a series of sinusoidal intensity patterns 
\begin{equation}\label{C04}
	s\left({r}_\text{o},\alpha, \beta, p\right)=1-\bar{\mathbf{m}}\cos \left (2\pi {p_{\alpha}}\cdot {r_\text{o}}+\beta \right ),
\end{equation}
where $\alpha \in \text{A}= \left [ 0, \pi/3,2\pi/3\right ]$ indicates the orientation of sinusoidal illumination pattern, $\beta\in \text{B} = \left [ 0, 2\pi/3,4\pi/3\right ]$ is the phase of illumination pattern, $\bar{\mathbf{m}}=0.5$ is the modulation factor, ${p_{\alpha }}=( p\cos\alpha, p\sin\alpha)$ is sinusoidal illumination frequency vector in reciprocal space, and ${{r}_\text{o}} = ( {{x_\text{o}},{y_\text{o}}} )$ is the coordinates of object plane, respectively. Generally speaking, the maximum of the amplitude of sinusoidal frequency vector $p$ is $\frac{1}{\Delta x_R}$, determined by the spatial resolution of the imaging setup. Moreover, $s\left({r}_\text{o},\alpha, \beta,p\right)$ is also composed of $m\times n$ pixels which can create a $K\times K$ diagonal matrix vector
\begin{equation}\label{C05}
	\text{S}\left(\alpha, \beta,p\right)=\left[ \begin{array}{l}
		{s\left(1,1, \alpha, \beta,p \right)},0, \cdots ,{\kern 1pt} 0\,\\
		0\,,{s\left(1,2, \alpha, \beta,p \right)}, \cdots ,0\\
		\; \vdots \quad \;\;\;\;\; \ddots \;\;\quad\quad\quad \;\;\;{\kern 1pt} \; \vdots \\
		0,0, \cdots ,{s\left(m,n, \alpha, \beta,p \right)}
	\end{array} \right].
\end{equation}
After the speckle modulated by the sinusoidal intensity patterns $s\left({r}_\text{o},\alpha, \beta,p\right)$, the matrix operations of raw-image of SIGI can be expressed as 
\begin{equation}\label{C06}
	sigi\left(\alpha, \beta,p\right)= \frac{1}{N} {\Psi^\text{T}}\text{M}\left(\alpha, \beta,p\right) O,
\end{equation}
where $\text{M}\left(\alpha, \beta,p\right) = \Phi \text{S}\left(\alpha, \beta,p\right)$ is the new measurement matrix after the speckle modulated by the sinusoidal patterns $s\left({r}_\text{o},\alpha, \beta, p\right)$. Comparing with the traditional GI by Eq.~(\ref{C03}), the only difference of matrix operation by Eq.~(\ref{C06}) is that the sinusoidal structured speckle replaces the original speckle in the bucket detector signal arm. Theoretically, those sinusoidal structured speckle can be created by the free diffraction of computer-generated holography (CGH). At present, CGH can be created by the Gale-Shapley (GS) algorithm~\cite{Gerchberg1972Apractical} and related improved algorithms~\cite{Liu2022Double,Leonardo2007Computer,Akahori1986Spectrum}. In fact, if a charge coupled device (CCD) acts as the bucket detector in the SIGI scheme~\cite{Li2019PRA,Li2021Eigenmode}, this sinusoidal modulated speckle field can be achieve by a subsequent data processing. According to the reconstruction algorithm of structured illumination imaging~\cite{Lal2016Structured}, the SIGI with modulated illumination frequency vector $p$ can be create by
\begin{equation}\label{C07}
	SIGI(p)=\mathop \Upsilon \limits_{\left( {\alpha, \beta } \right) \in \Omega } \{ {sigi\left( {\alpha, \beta,p } \right)} \},
\end{equation}
where the operator $\Upsilon$ means the spectrum merge by 9 raw-images $sigi\left(\alpha, \beta,p\right)$, and the modulation set $\Omega$ originates from the Cartesian product $\text{A}\times \text{B}=\left\{(\alpha, \beta)|\alpha\in \text{A}, \beta\in \text{B} \right\}$. In the Secs.~\ref{EandNS} and~\ref{Discussion}, we averaged spectrum originated from those saw-images at overlapping region to build the SIGI. The detailed description on spectrum merge can refer to the reference~\cite{Lal2016Structured}.

In some GI setups, there is a distance between the object and the bucket detector as shown by the inset in Fig.~\ref{SIGIdiagram}. At this moment, the bucket detector just collects the diffraction signal of the transmitted speckle from the test object. Therefore, the matrix format of diffractive GI can be expressed:
\begin{equation}\label{C08}
	GI_{\text{D}} = \frac{1}{N}{\Psi^\text{T}}\bigtriangleup\{\Phi O\},
\end{equation}
where the operator $\bigtriangleup\{\cdot\}$ means the free diffraction and $\bigtriangleup\{\Phi O\}_{(N\times 1)}$ is bucket signal in the diffractive GI scheme. Similarly, the diffractive SIGI can be create by
\begin{equation}\label{C09}
	SIGI_\text{D}(p)=\mathop \Upsilon \limits_{\left( {\alpha, \beta } \right) \in \Omega } \{ {sigi_\text{D}\left( {\alpha, \beta,p } \right)} \},
\end{equation}
where
\begin{equation}\label{C010}
	sigi_\text{D}\left(\alpha, \beta,p\right)= \frac{1}{N} {\Psi^\text{T}} \bigtriangleup\{\text{M}\left(\alpha, \beta,p\right) O\}.
\end{equation}
In the following section, we will verify the spatial high-resolution of those two SIGI schemes.
\section{Numerical Experiment} \label{EandNS}
\begin{figure}[t]
	\centering
	\includegraphics[width=0.49\textwidth]{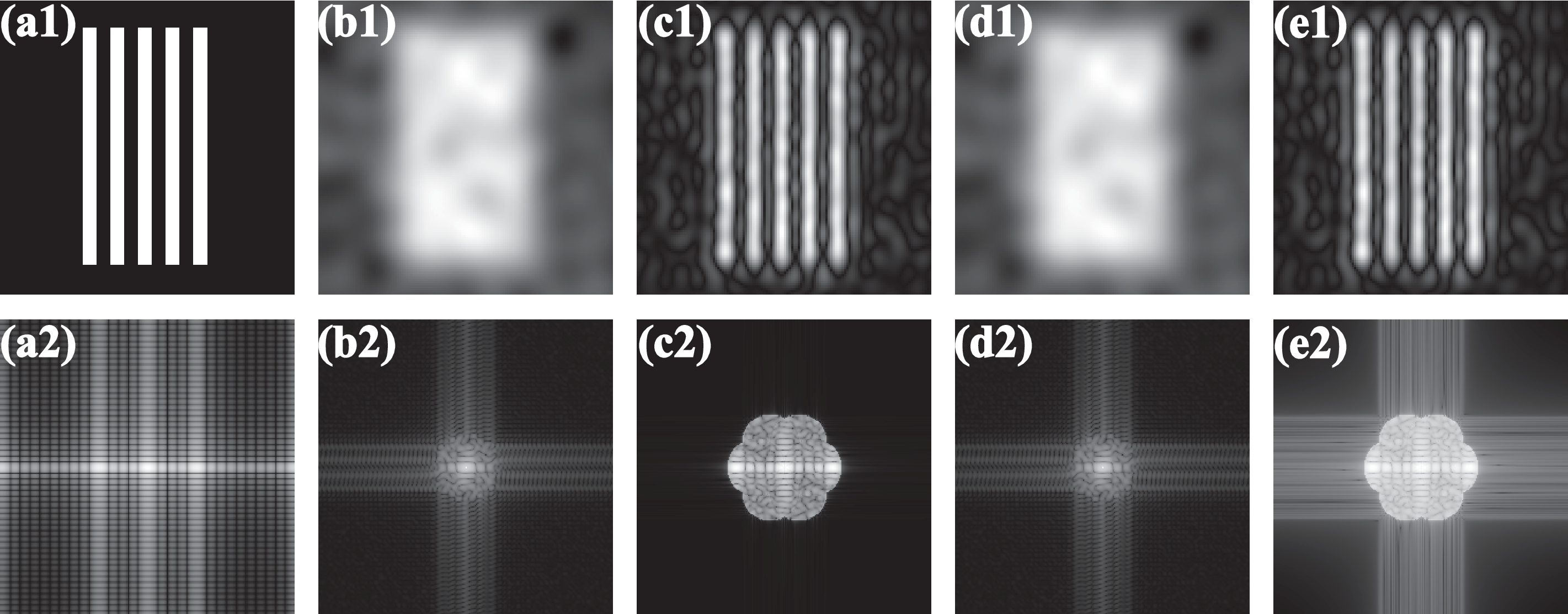}
	\caption{(a1) Test object. (b1) Non-diffractive GI. (c1) Non-diffractive SIGI. (d1) Diffractive GI. (e1) Diffractive SIGI. (a2-e2) Logarithmic transformation of Fourier domain of images in (a1-e1), respectively.} \label{GIandSIGI}
\end{figure}

To test the high-resolution of SIGI, we perform the follow-up numerical experiments. In those SIGI schemes, the sinusoidal structured speckle $\text{M}\left(\alpha, \beta,p\right)$ was not produced by holography, but fabricated by the multiplication of the speckle and correspondingly sinusoidal intensity patterns. In order to generate the dynamic speckle illumination, the phase structure $e^{i\theta (r_\text{s},t)-ikr_{\text{s}}^{2}/(2z)}$ were loaded on the SLM, where the phasor $\theta (r_\text{s},t)$ is completely random in the time and spatial distribution. According to calculation, the Rayleigh-resolution limit $\Delta x_R$ of the non-diffractive GI scheme is 433 $\mu \text{m}$, where the bucket detector was placed as close as possible to the object. For the diffractive GI scheme, the diffraction distance $z_0$ is 0.5 m between the bucket detector and the object. In our numerical experiment, we added up all the light intensity on the plane of bucket detector to ensure that all transmitted light was collected.

We choose a 0-1 resolution test object with 128×128 pixels, which consists alternative opaque and clear stripes of equal width as shown in Fig.~\ref{GIandSIGI}(a1). Here the number of total frames $N$ is 5000 for GI and SIGI schemes in Fig.~\ref{GIandSIGI}. As expected, stripes of the test object are indistinguishable in the non-diffractive GI scheme as shown in Fig.~\ref{GIandSIGI}(b1), due to the spatial resolution of non-diffractive GI scheme much larger than the spacing 250 $\mu \text{m}$ of adjacent fringe in the test object. On the other hand, all stripes in test object are  distinguishable in the SIGI scheme as shown in Fig.~\ref{GIandSIGI}(c1). In theory~\cite{Lal2016Structured}, the spatial resolution of SIGI scheme is $\Delta x_R/2=216.5$ $ \mu \text{m}$. In addition, compared with the diffractive GI as shown in Fig.~\ref{GIandSIGI}(d1), the diffractive SIGI scheme also improves the imaging resolution as shown in Fig.~\ref{GIandSIGI}(e1). To illustrate the extended-spectrum by structured illumination, figures~\ref{GIandSIGI}(a2-e2) present the logarithmic transformation $\text{ln}\left ( 1+\left | \mathscr{F}\left ( \text{image} \right )  \right |  \right ) $ of Fourier domain of images in figures~\ref{GIandSIGI}(a1-e1), respectively. It can be seen that the radius of Fourier domain of SIGI is nearly twice as much as the corresponding GI. 

The fidelity of imaging can be estimated by calculating the peak signal-to-noise ratio (PSNR):
\begin{equation}\label{C11}
	\begin{split}
		PSNR = 10 \log_{10} {\left[ {{( {2^{\text{B}} - 1} )}^2}/MSE \right]},
	\end{split}
\end{equation}
where $\text{B}=8$ is the bit depth of images and $MSE$ is the mean-square error of the result with respect to the test object. Here, $MSE$ is defined as:
\begin{equation}\label{C12}
	\begin{split}
		MSE=\frac{1}{\tau^2} \sum\limits_{m=1}^\tau\sum\limits_{n=1}^\tau {\left[GI(m,n)-o(m,n) \right]^2},
	\end{split}
\end{equation}
where the result of those ghost images $GI(m,n)$ reshaped to two-dimension and the test object $o(m,n)$ are digitized to $0\sim2^{\text{B}}-1$, $\tau=128$ is the total number of pixels in single dimension, respectively. It can be seen that the higher the quality of the result, the greater the PSNR. According to calculation, those PSNR of Figs.~\ref{GIandSIGI}(b1-e1) are 7.64 dB, 14.56 dB, 7.64 dB and 14.22 dB, respectively. In addition, the structural similarity (SSIM)~\cite{Wang2004Image} are also used as our evaluation index to quantitatively analyze the quality of those ghost images at different schemes. The SSIM value of Figs.~\ref{GIandSIGI}(b1-e1) are 0.015, 0.307, 0.015 and 0.298, respectively.
It can be seen that the quality of ghost image results has been optimized through the structured speckle illumination.

\section{Discussion} \label{Discussion}
\subsection{the structured illumination ghost imaging with pseudo-modulation} \label{Discussion_sub01}
\begin{figure}[b]
	\centering
	\includegraphics[width=0.49\textwidth]{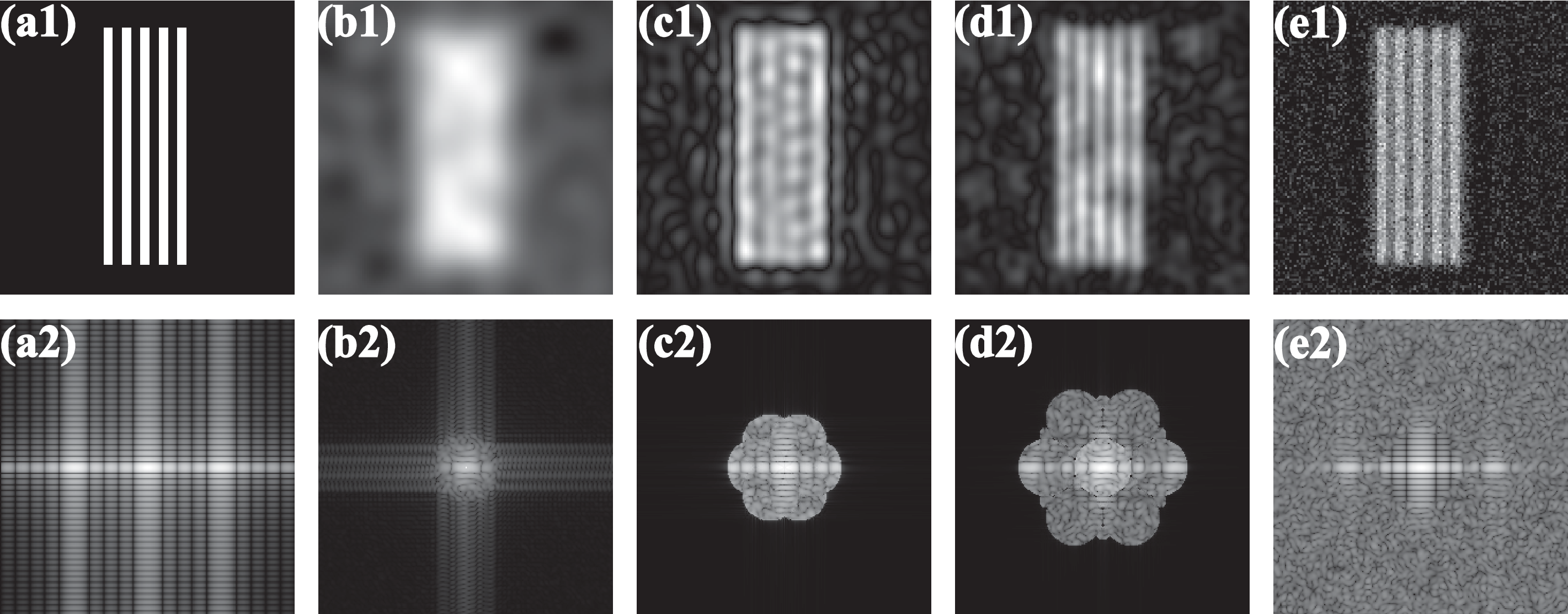}
	\caption{(a1) Test object. (b1) GI. (c1) SIGI. (d1) Pseudo-modulation SIGI. (e1) PGI. (a2-e2) Logarithmic transformation of Fourier domain of images in (a1-e1), respectively.} \label{SIGIandPGI}
\end{figure}
Actually, besides authentic speckle modulation by theoretical CGH, a pseudo-modulation of speckle structured illumination with frequency vector $p'(>p)$ is effective for SIGI, which can be available by
\begin{equation}\label{C13}
	SI_{\text{pse}}GI(p')=\mathop \Upsilon \limits_{\left( {\alpha, \beta } \right) \in \Omega '} \{ {si_{\text{pse}}gi\left( {\alpha, \beta,p' } \right)} \},
\end{equation}
where 
\begin{equation}\label{C14}
	si_{\text{pse}}gi\left(\alpha, \beta,p'\right)= \frac{1}{N} {\Psi^\text{T}}\text{M}\left(\alpha, \beta\right){\Phi^{-1}} \Phi O.
\end{equation}
Here, $\Phi^{-1}$ is the Moore–Penrose pseudo-inverse of the matrix $\Phi$. Interestingly, if the transmission coefficient $\Phi O$ is available in the traditional CGI, $si_{\text{pse}}gi\left(\alpha, \beta,p'\right)$ originated from structural speckle illumination with arbitrary modulation frequency $p'$ and arbitrary modulation orientation $\left(\alpha, \beta\right)$ can be achieved. Thus, $SI_{\text{pse}}GI$ can break the upper limit $\frac{1}{\Delta x_R}$ of modulation frequency $p$ in the SIGI with authentic speckle modulation. Therefore, the spatial resolution of pseudo-modulation SIGI scheme will be better than SIGI scheme in theory. For the pseudo-modulation SIGI, we merge the spectrum of three sets of structural illumination $\left( {{\rm{A'}} \times {\rm{B}},2p/\sqrt 3 } \right)$, $\left( {{\rm{A}} \times {\rm{B}},2p} \right)$, and $\left( {{\rm{A'}} \times {\rm{B}},4p/\sqrt 3 } \right)$, respectively. Here, a new orientation of sinusoidal illumination pattern ${{\rm{A'}}}$ is $\left [\pi/6,3\pi/6,5\pi/6\right]$. Actually, the pseudo-modulation SIGI scheme has some similarities with the PGI scheme. Comparing with the imaging algorithm $PGI=\frac{1}{N}\Psi^\text{-1}\Psi O$~\cite{Gong2015Highpr}, we find the imaging quality are almost identical with the modified algorithm $\frac{1}{N}\Phi^\text{-1}\Phi O$. Therefore, the matrix operation of Eq.~(\ref{C14}), namely the raw image of pseudo-modulation SIGI, can be rewritten as 
\begin{equation}\label{C1402}
	si_{\text{pse}}gi\left(\alpha, \beta,p'\right)=  {\Psi^\text{T}}\text{M}\left(\alpha, \beta\right)PGI.
\end{equation}
It's clear that the spectrum domain of raw image by the pseudo-modulation SIGI is only a subset of PGI's. What's more, ghost image established by pseudo-modulation SIGI undergos the spectrum filtering by the recovery algorithm of structured illumination~\cite{Lal2016Structured}. Interestingly, the pseudo-modulation SIGI demonstrates the sub-Rayleigh spatial resolution of PGI scheme.

Figure~\ref{SIGIandPGI} shows the numerical simulation demonstration of spatial resolution superiority of the pseudo-modulation SIGI. Here the number of total frames $N$ is 5000 for those GI schemes. Figure~\ref{SIGIandPGI}(a1) show the test object with 128×128 pixels and the spacing 167 $\mu \text{m}$ of adjacent fringe. Using the same imaging methods in Fig.~\ref{GIandSIGI}(b1) and Fig.~\ref{GIandSIGI}(c1), the imaging results of GI and SIGI are shown in the Fig.~\ref{SIGIandPGI}(b1) and Fig.~\ref{SIGIandPGI}(c1), respectively. Due to the further refinement of the object's fringe, GI and SIGI only show a rough outline of the test object. Figure~\ref{SIGIandPGI}(d1) and figure~\ref{SIGIandPGI}(e1) show the pseudo-modulation SIGI and the PGI, respectively. Although both of them distinguish the stripes of the object, they have their own characteristics. Compared with PGI scheme, pseudo-modulation SIGI shows a slightly lower spatial resolution but gives a smoother image. Figures~\ref{SIGIandPGI}(a2-e2) present the logarithmic transformation $\text{ln}\left ( 1+\left | \mathscr{F}\left ( \text{image} \right )  \right |  \right ) $ of Fourier domain of images in figures~\ref{SIGIandPGI}(a1-e1), respectively. Compared with the maximum frequency limit $2p$ in SIGI as shown in the Fig.~\ref{SIGIandPGI}(c2), the maximum frequency of the pseudo-modulation SIGI here is almost $3p$ as shown in the Fig.~\ref{SIGIandPGI}(d2). As a result, the pseudo-modulation SIGI has significantly better spatial resolution than SIGI. 

According to calculation, those PSNR of Figs.~\ref{SIGIandPGI}(b1-e1) are 8.90 dB, 11.60 dB, 13.46 dB and 13.91 dB, respectively. In addition, the SSIM value of Figs.~\ref{SIGIandPGI}(b1-e1) are 0.012, 0.068, 0.158 and 0.164, respectively. It can be seen that the quality of ghost image results has been optimized through the pseudo-modulation speckle illumination. Although the spatial resolution of PGI scheme is better than other schemes, there is inevitable white noise in the high-frequency originated from the searching algorithm of the Moore–Penrose pseudo-inverse. Therefore, the extended spectrum of low-frequency of PGI by the structured illumination will be beneficial for the ghost image quality. This will be discussed in the following section in detail.

\subsection{the structured illumination ghost imaging grafted with the pseudo-inverse ghost imaging} \label{Discussion_sub02}
To further optimize ghost image, two types of sub-Rayleigh schemes the SIGI and the PGI can be combined to create the structured illumination pseudo-inverse ghost imaging (SIPGI), which is available by \begin{equation}\label{C15}
	SIPGI(p)=\mathop \Upsilon \limits_{\left( {\alpha, \beta } \right) \in \Omega } \{ sipgi\left( {\alpha, \beta,p } \right) \},
\end{equation}
where
\begin{equation}\label{C16}
	sipgi\left(\alpha, \beta,p\right)= \frac{1}{N} {\Psi^\text{-1}}\text{M}\left(\alpha, \beta,p\right) O.
\end{equation}
What's more, a pseudo-modulation SIPGI scheme with three sets of structural illumination $\left( {{\rm{A'}} \times {\rm{B}},2p/\sqrt 3 } \right)$, $\left( {{\rm{A}} \times {\rm{B}},2p} \right)$, and $\left( {{\rm{A'}} \times {\rm{B}},4p/\sqrt 3 } \right)$ can be available by
\begin{equation}\label{C17}
	SI_{\text{pse}}PGI(p')=\mathop \Upsilon \limits_{\left( {\alpha, \beta } \right) \in \Omega '} \{ si_{\text{pse}}pgi\left( {\alpha, \beta,p' } \right) \},
\end{equation}
where
\begin{equation}\label{C18}
	si_{\text{pse}}pgi\left(\alpha, \beta,p'\right)= \frac{1}{N} {\Psi^\text{-1}}\text{M}\left(\alpha, \beta,p'\right){\Phi^{-1}} \Phi O.
\end{equation}
\begin{figure}[t]
	\centering
	\includegraphics[width=0.49\textwidth]{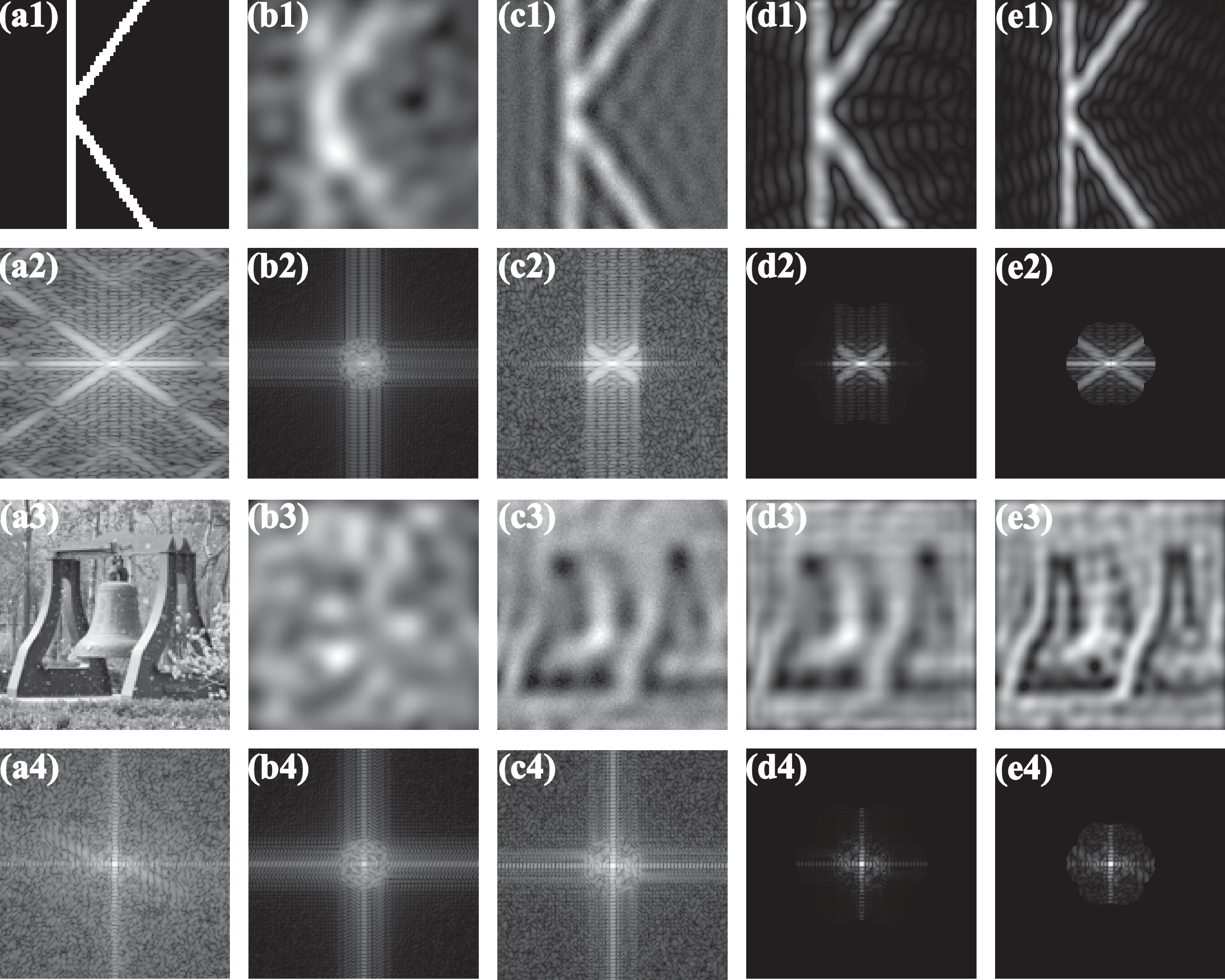}
	\caption{Test objects:(a1),(a3); GI:(b1),(b3); PGI:(c1),(c3); Pseudo-modulation SIPGI:(d1),(d3); SIPGI:(e1),(e3); (a2-e2, a4-e4) Logarithmic transformation of Fourier domain of images in (a1-e1, a3-e3), respectively.} \label{SIpsePGIandSIPGI}
\end{figure}
Figure~\ref{SIpsePGIandSIPGI} gives the numerical experiment with a 0-1 binary object 'K' and a natural grayscale object NANKAI Bell. The number of total frames $N$ is 500. Figures~\ref{SIpsePGIandSIPGI} (a1,a3) show those test objects, which are 128×128 pixels. Using the same imaging methods in Fig.~\ref{SIGIandPGI}(b1) and Fig.~\ref{SIGIandPGI}(e1), the imaging results of GI and PGI are shown in the Figs.~\ref{SIpsePGIandSIPGI}(b1,b3) and Figs.~\ref{SIpsePGIandSIPGI}(c1,c3), respectively. Compared with the GI, the PGI scheme can produces an better imaging quality. What's more, the imaging results of the pseudo-modulation SIPGI and the authentic-modulation SIPGI are shown in the Figs.~\ref{SIpsePGIandSIPGI}(d1,d3) and Figs.~\ref{SIpsePGIandSIPGI}(e1,e3), respectively. According to calculation, those PSNRs of Figs.~\ref{SIpsePGIandSIPGI}(b1-e1) with the 0-1 binary object are 9.33 dB, 11.14 dB, 14.91 dB and 16.54 dB, respectively. In addition, the SSIM value of Figs.~\ref{SIpsePGIandSIPGI}(b1-e1) are 0.014, 0.039, 0.127 and 0.230, respectively. What's more, those PSNRs of Figs.~\ref{SIpsePGIandSIPGI}(b3-e3) with the garyscale object are 12.48 dB, 16.73 dB, 16.38 dB and 17.52 dB, respectively. The SSIM value of Figs.~\ref{SIpsePGIandSIPGI}(b3-e3) are 0.077, 0.242,  0.229 and 0.376, respectively. It can be seen that the quality of ghost image results has been optimized through the authentic-modulation SIPGI scheme.

In fact, the advantages and disadvantages of imaging quality of these results can be intuitively displayed in the spectrum domain. Figures~\ref{SIpsePGIandSIPGI}(a2-e2,a4-e4) present the logarithmic transformation $\text{ln}\left ( 1+\left | \mathscr{F}\left ( \text{image} \right )  \right |  \right ) $ of Fourier domain of images in figures~\ref{SIpsePGIandSIPGI} (a1-e1,a3-e3), respectively. Compared with the PGI, the pseudo-modulation SIPGI scheme does not realize broaden the detected spectrum domain of imaging system but the authentic-modulation SIPGI scheme achieves the goal. For the pseudo modulation SIPGI, the detected spectrum as shown by the Figs.~\ref{SIpsePGIandSIPGI}(d2,d4) can not be shifted with pseudo modulation speckle illumination. However, the detected spectrum as shown by the Figs.~\ref{SIpsePGIandSIPGI} (e2,e4) can be shifted in the authentic-modulation SIPGI scheme and result to an better image quality than other GI schemes. Here, we reiterate once again that the authentic-modulation of the sinusoidal structured speckle illumination $\text{M}\left(\alpha, \beta,p\right)$ is not produced by GCH, but fabricates by the matrix multiplication. It can be seen that the realization of the sinusoidal structured speckle illumination by GCH is very meaningful for the sub-Rayleigh ghost imaging. This will be our future research plan.

\section{Conclusion} \label{Conclusion}
In conclusion, we proposed theoretically and demonstrated numerical experimentally the sub-Rayleigh SIGI scheme. In addition, a pseudo-modulation SIGI scheme, whose modulation frequency breaks the upper limit of imaging system, can achieve a better imaging resolution. Finally, combined the SIGI scheme and the PGI scheme, we propose the SIPGI scheme which can greatly enhance the ghost imaging quality whether the 0-1 binary object or a natural grayscale object.


\section*{Funding.} National Natural Science Foundation of China (NSFC) (62105188).

\section*{Acknowledgements.} Liming Li thanks Dezhong Cao (Yantai University) for helpful discussions.




\begin{thebibliography}{10}
	\bibitem{Rayleigh1897Investigations}
	F. R. S. Lord Rayleigh, "XXXI. Investigations in optics, with special reference to the spectroscope," Phil. Mag. {\bfseries 8}, 261-274 (1879).
	
	\bibitem{Goodman2005Introduction2}
	J. W. Goodman, {\it Introduction to Fourier Optics} (Robert \& Company, 2005).
	
	\bibitem{Heintzmann1998Laterally}
	R. Heintzmann and C. G. Cremer, "Laterally modulated excitation microscopy: improvement of resolution by using a diffraction grating," Proc. SPIE {\bfseries 3568}, 185-196 (1999).
	
	\bibitem{Gustafsson2001Microscopy}
	M. G. L. Gustafsson, "Surpassing the lateral resolution limit by a factor of two using structured illumination microscopy," J. Microsc. {\bfseries 198}, 82-87 (2000).
	
	\bibitem{Frohn2000PNAS}
	J. T. Frohn, H. F. Knapp, and A. Stemmer, "True optical resolution beyond the Rayleigh limit achieved by standing wave illumination," Proc. Natl. Acad. Sci. USA {\bfseries 97}, 7232-7236 (2000).
	\bibitem{Heintzmann2017Super}
	R. Heintzmann, and T. Huser, "Super-resolution structured illumination microscopy," Chem. Rev. {\bfseries 117}, 13890-13908 (2017).
	
	\bibitem{Strohl2016Frontiers}
	F. Str\"{o}hl and C. F. Kaminski, "Frontiers in structured illumination microscopy," Optica {\bfseries 3}, 667-677 (2016).
	
	\bibitem{Xue2018Three}
	Y. Xue and Peter T. C. So, "Three-dimensional super-resolution high-throughput imaging by structured illumination STED microscopy," Opt. Express {\bfseries 26}, 20920-20928 (2018).
	
	\bibitem{Wang2023Hybrid}
	J. Wang, J. Fan, B. Zhou, \emph{et al.}, "Hybrid reconstruction of the physical model with the deep learning that improves structured illumination microscopy," Adv. Photon. Nexus {\bfseries 2}, 016012-016012 (2023).
	
	\bibitem{Chen2023Superresolution}
	X. Chen, S. Zhong, Y. Hou, \emph{et al.}, "Superresolution structured illumination microscopy reconstruction algorithms: a review," Light Sci. Appl. {\bfseries 12}, 172 (2023).
	\bibitem{Bennink2002CoincidencePRL}
	R. S. Bennink, S. J. Bentley, and R. W. Boyd,""Two-photon" coincidence imaging with a classical source," Phys. Rev. Lett. {\bfseries 89}, 113601 (2002).
	
	\bibitem{Valencia2005TwoPRL}
	A. Valencia, G. Scarcelli, M. D’Angelo, \emph{et al.}, " Phys. Rev. Lett. {\bfseries 94}, 063601 (2005).
	
	\bibitem{Ferri2005HighPRL}
	F. Ferri, D. Magatti, A. Gatti, \emph{et al.}, "High-resolution ghost image and ghost diffraction experiments with thermal light," Phys. Rev. Lett. {\bfseries 94}, 183602 (2005).
	
	\bibitem{ChenHigh-visibility2010}
	X.-H. Chen, I. N. Agafonov, K.-H. Luo, \emph{et al.}, "High-visibility, high-order lensless ghost imaging with thermal light," Opt. Lett. {\bfseries 35}, 1166-1168 (2010).
	
	\bibitem{Wu2005Correlated}
	D. Zhang, Y.-H. Zhai, L.-A. Wu, \emph{et al.}, "Correlated two-photon imaging with true thermal light," Opt. Lett. {\bfseries 30}, 2354-2356 (2005).
	\bibitem{Liu2014Lensless}
	X.-F. Liu, X.-H. Chen, X.-R. Yao, \emph{et al.}, "Lensless ghost imaging with sunlight," Opt. Lett. {\bfseries 39}, 2314-2317 (2014).
	
	\bibitem{Klyshko1988Two}
	D.N. Klyshko, "Two-photon light: influence of filtration and a new possible EPR experiment," Phys. Lett. A {\bfseries 128}, 133-137 (1988).
	
	\bibitem{Pittman1995PRA}
	T. B. Pittman, Y. H. Shih, D. V. Strekalov, \emph{et al.}, "Optical imaging by means of two-photon quantum entanglement," Phys. Rev. A {\bfseries 52}, R3429 (1995).
	
	\bibitem{Abouraddy2004Entangled}
	A. F. Abouraddy, P. R. Stone, A. V. Sergienko, \emph{et al.}, "Entangled-photon imaging of a pure phase object," Phys. Rev. Lett. {\bfseries 93}, 213903 (2004).
	
	\bibitem{Aspden2015Photon}
	R. S. Aspden, N. R. Gemmell, P. A. Morris, , \emph{et al.}, "Photon-sparse microscopy: visible light imaging using infrared illumination,"  Optica {\bfseries 2}, 1049-1052 (2015).
	\bibitem{Shapiro2008Computational}
	J. H. Shapiro, "Computational ghost imaging," Phys. Rev. A {\bfseries 78}, 061802(R) (2008).
	
	\bibitem{Bromberg2009Ghost}
	Y. Bromberg, O. Katz, and Y. Silberberg, "Ghost imaging with a single detector," Phys. Rev. A {\bfseries 79}, 053840 (2009).
	
	\bibitem{Sun2013science}
	B. Sun, M. P. Edgar, R. Bowman, \emph{et al.}, "3D computational imaging with single-pixel detectors," Science {\bfseries 340}, 844-847 (2013).
	
	\bibitem{XU20181000}
	Z.-H. Xu, W. Chen, J. Penuelas, \emph{et al.}, "1000 fps computational ghost imaging using LED-based structured illumination," Opt. Express {\bfseries 26}, 2427-2434 (2018).
	
	\bibitem{Edgar2018Principles}
	M. P. Edgar, G. M. Gibson, and M. J. Padgett, "Principles and prospects for single-pixel imaging," Nat. Photonics {\bfseries 13}, 13-20 (2019).
	\bibitem{Sun2019Single}
	M.-J. Sun and J.-M. Zhang, "Single-pixel imaging and its application in three-dimensional reconstruction: a brief review," Sensors {\bfseries 19}, 732 (2019).
	
	\bibitem{Hong2019Sub}
	P. Hong, "Sub-Rayleigh single-pixel imaging via optical fluctuation," Opt. Lett. {\bfseries 44}, 1754-1757 (2019).
	
	\bibitem{Li2019PRA}
	L. Li, P. Hong, and G. Zhang, "Transverse revival and fractional revival of the Hanbury Brown and Twiss bunching effect with discrete chaotic light," Phys. Rev. A {\bfseries 99}, 023848 (2019).
	
	\bibitem{Li2021Eigenmode}
	L. Li, B. Wang, S. Li, \emph{et al.}, "Eigenmode high-visibility imaging in the far-field ghost imaging system with discrete chaotic light," Phys. Lett. A {\bfseries 420}, 127749 (2021).
	
	\bibitem{Kner2009Super}
	P. Kner, B. B Chhun, E. R Griffis, \emph{et al.}, "Super-resolution video microscopy of live cells by structured illumination," Nat. Methods {\bfseries 6}, 339-342 (2009).
	\bibitem{Hirvonen2009Structured}
	L. M. Hirvonen, K. Wicker, O. Mandula, \emph{et al.}, "Structured illumination microscopy of a living cell," Eur. Biophys. J. {\bfseries 38}, 807-812 (2009).
	
	\bibitem{Chang2009Isotropic}
	B.-J. Chang, L.-J. Chou, Y.-C. Chang, \emph{et al.}, "Isotropic image in structured illumination microscopy patterned with a spatial light modulator," Opt. Express {\bfseries 17}, 14710-14721 (2009).
	
	\bibitem{Guo2018Visualizing}
	Y. Guo, D. Li, S. Zhang, \emph{et al.}, "Visualizing intracellular organelle and cytoskeletal interactions at nanoscale resolution on millisecond timescales," Cell {\bfseries 175}, 1430-1442 (2018).
	
	\bibitem{Gong2015Highsr}
	W. Gong, and S. Han, "High-resolution far-field ghost imaging via sparsity constraint," Sci. Reports, {\bfseries 5}, 9280 (2015).
	
	\bibitem{Lochocki2021Ultimate}
	B. Lochocki, K. Abrashitova, J. F. de Boer, \emph{et al.}, "Ultimate resolution limits of speckle-based compressive imaging," Opt. Express {\bfseries 29}, 3943-3955 (2021).
	\bibitem{Zhang2014Object}
	C. Zhang, S. Guo, J. Cao, \emph{et al.}, "Object reconstitution using pseudo-inverse for ghost imaging," Opt. Express {\bfseries 22}, 30063-30073 (2014).
	
	\bibitem{Gong2015Highpr}
	W. Gong, "High-resolution pseudo-inverse ghost imaging," Photon. Res. {\bfseries 3}, 234-237 (2015).
	
	\bibitem{Zhang2016High}
	S. Zhang, W. Wang, R. Yu, \emph{et al.}, "High-order correlation of non-Rayleigh speckle fields and its application in super-resolution imaging," Laser Phys. {\bfseries 26}, 055007 (2016).
	
	\bibitem{KYRUS2016High}
	K. Kuplicki and K. W. C. Chan, "High-order ghost imaging using non-Rayleigh speckle sources," Opt. Express {\bfseries 24}, 26766-26776 (2016).
	
	\bibitem{Sprigg2016Super}
	J. Sprigg, T. Peng, and Y. Shih, "Super-resolution imaging using the spatial-frequency filtered intensity fluctuation correlation," Sci. Rep. {\bfseries 6}, 38077 (2016).
	\bibitem{Meng2018Super}
	S.-Y. Meng, Y.-H. Sha, Q. Fu, \emph{et al.}, "Super-resolution imaging by anticorrelation of optical intensities," Opt. Lett. {\bfseries 43}, 4759 (2018).
	
	\bibitem{Wang2022Far}
	F. Wang, C. Wang, M. Chen, \emph{et al.}, "Far-field super-resolution ghost imaging with a deep neural network constraint," Light Sci. Appl. {\bfseries 11}, 1 (2022).
	
	\bibitem{Katz2009Compressive}
	O. Katz, Y. Bromberg, and Y. Silberberg, "Compressive ghost imaging," Appl. Phys. Lett. {\bfseries 95}, 131110 (2009).
	
	\bibitem{Gerchberg1972Apractical}
	R. W. Gerchberg and W. O. Saxton, "A practical algorithm for the determination of plane from image and diffraction pictures," Optik {\bfseries 35}, 237-246 (1972).
	
	\bibitem{Akahori1986Spectrum}
	H. Akahori, "Spectrum leveling by an iterative algorithm with a dummy area for synthesizing the kinoform," Appl. Opt. {\bfseries 25}, 802-811 (1986).
	\bibitem{Leonardo2007Computer}
	R. Di Leonardo, F. Ianni, and G. Ruocco, "Computer generation of optimal holograms for optical trap arrays," Opt. Express {\bfseries 15}, 1913-1922 (2007).
	
	\bibitem{Liu2022Double}
	K. Liu, Z. He, and L. Cao, "Double amplitude freedom Gerchberg–Saxton algorithm for generation of phase-only hologram with speckle suppression," Appl. Phys. Lett. {\bfseries 120}, 061103 (2022).
	
	\bibitem{Lal2016Structured}
	A. Lal, C. Shan, and P. Xi, "Structured Illumination Microscopy Image Reconstruction Algorithm," IEEE J. Sel. Top. Quantum Electron. {\bfseries 22}, 50-63 (2016).
	
	\bibitem{Wang2004Image}
	Z. Wang, A.C. Bovik, H.R. Sheikh, \emph{et al.}, "Image quality assessment: from error visibility to structural similarity," IEEE TIP. {\bfseries 13}, 600-612 (2004).
	
\end{thebibliography}
\end{document}